\documentstyle[11pt]{article}

\def\eps{\epsilon}
\def\sp{\phantom{s}}
\def\dual#1{{{^*}#1}}
\def\ph{\phantom}
 
\def\pa{\partial}
\def\k{\kappa}
 \def\G{\Gamma}
\def\a{\alpha}
\def\b{\beta}
\def\d{\delta} 
\def\e{\epsilon}

\def\k{\kappa}
\def\l{\lambda} 
\def\m\mu
\def\n{\nu}

\def\eps{\epsilon}

\def\be{\begin{equation}}
\def\ee{\end{equation}}
\def\mn{{\mu \nu}}
\def\ab{{\alpha\beta}}

\begin{document}

\title{Tree Amplitudes and Two-loop Counterterms \\
in D=11 Supergravity}
\author{S. Deser$^{a}$ and D. Seminara$^{b}$\thanks{CEC Post-doctoral
Fellow under contract FMRX CT 96-0045.}\\ \it $^{a}$Physics
Department, Brandeis University, Waltham, MA 02454, USA\\ \it
$^{b}$Laboratoire de Physique Th{\'e}orique\thanks{Unit{\'e} Mixte UMR
8549
associ{\'e}e au  Centre de la Recherche Scientifique et {\`a} l'
{\'E}cole Normale Sup{\'e}rieure.}, {\'E}cole Normale
Sup{\'e}rieure\\ \it F-75231, Paris CEDEX 05, France}
\maketitle

\begin{abstract} We compute the tree level  4-particle bosonic
scattering amplitudes in D=11 supergravity.  By construction, they
are part of a linearized supersymmetry-, coordinate- and 3-form
gauge-invariant. While this on-shell invariant is nonlocal,
suitable SUSY-preserving differentiations turn it into a local one
with correct dimension to provide a natural lowest (two-loop)
order counterterm candidate. Its existence shows explicitly that
no symmetries protect this ultimate supergravity from the
nonrenormalizability of its lower-dimensional counterparts.
\end{abstract}

\vfill

\begin{flushright} LPT-ENS-99/52\\ BRX-TH-456\\
\end{flushright}
\vfill

\section{Introduction}

In the post-D=10 superstring era, D=11 supergravity (SUGRA)
\cite{cremmer} has again attracted the attention it has always
deserved, without however, becoming any easier to handle
technically. In particular, supersymmetry (SUSY) invariants are
still (absent an appropriate calculus) difficult to verify, let
alone construct. Here, we will supply (the linearized bosonic part
of) one such invariant. Our work had two motivations beyond its
intrinsic interest within the theory. Most directly, we wanted to
determine unambiguously whether there exist local invariants that
can serve as counterterms at lowest possible, here two-loop,
order. This nontrivial exercise has a historical basis in
lower-dimensional SUGRAs, where the existence of invariants is
easier to decide; there, no miracles occurred: counterterms were
always available. They sometimes started at higher order than in
pure Einstein gravity (GR) where every loop (except, accidentally,
one-loop at D=4) is dangerous. [For a recent historical review of
divergences in gravities see \cite{sdgermany}.] However, given all
the properties unique to D=11, and the fact that it is the last
frontier -- a local QFT that is non-ghost ({\it i.e.}, has no
quadratic curvature terms) and reduces to GR -- it is sufficiently
important not to give up hope before abandoning D=11 SUGRA (and
with it all QFTs incorporating GR) too quickly on
non-renormalizability grounds. Our second interest is in the
M-theoretical direction: any invariants that can be obtained here
might provide hints about the wider theory that presumably reduces
to D=11 SUGRA as its ``zero slope" limit.

The idea underlying our approach is that the set of all
$n$-particle (for fixed $n$) tree level scattering amplitudes
constructed within a perturbative expansion of the action is {\it
ipso facto} globally SUSY as well as linearized coordinate- and
3-form gauge invariant. Thus, because linearized SUSY does not mix
different powers of fields, the 4-point amplitudes of interest to
us, taken together, form an invariant. Also, within this lowest
order framework, the bosonic amplitudes are independent of
fermions: virtual ones cannot contribute at tree level. The above
statements together considerably lighten our task, which will be
to compute ``just" the parts involving the gravitational and form
bosonic excitations. The amplitudes involving fermions are not
necessarily more complicated, merely less relevant to our
immediate goal of reproducing terms about which the appropriate
divergence computations exist; indeed, we hope to return to them
\cite{waldron}. However, in order to use the scattering amplitudes
for counterterm purposes, it will first be necessary to strip them
of the nonlocality associated with exchange of the virtual
graviton and form particles (``formions") without compromising
their invariances. Actually, the task here will be not only to
remove nonlocality but to add sufficient further powers of
momentum to provide an on-shell invariant of correct dimension
that is an acceptable (and indeed first possible) perturbative
counterterm candidate. In this way, we will make contact with the
conclusive 2-loop results of \cite{bern}, where it was possible to
exhibit the infinity of a local 4-graviton term, one that is
precisely a component of our invariant. An earlier version of our
results was given in \cite{sdds}.

\section{Propagators and Vertices}

The basis for our computations is the full D=11 SUGRA action
\cite{cremmer}, expanded to quartic order in its two bosons,
namely the graviton and the formion, with three-form potential
$A_{\mu \nu\rho}$. The field strength $F_{\mu
\nu\alpha\beta}\equiv 4\partial_{[\mu }A_{\nu\alpha\beta]}$ is
invariant under the gauge transformations $\delta
A_{\nu\nu\alpha}=\partial_{[\mu } \xi_{\nu\alpha]}$, square
brackets denoting total (normalized) antisymmetrization. The
bosonic truncation of the Lagrangian is %
\begin{equation} \label{Lagra} %
{\cal  L}_B =-\frac{\sqrt{-g}}{4\kappa^2} R -\frac{\sqrt{-g}}{48}
F_{\mu \nu\rho\sigma}F^{\mu \nu\rho\sigma} +\frac{2 \kappa}{144^2}
\: \epsilon^{1 .. 11} A_{1-3}F_{4-7} F_{8-11}.
\end{equation}
The metric signature in eq. (\ref{Lagra}) is mostly minus, the
Ricci  tensor is defined by $R_{\alpha\beta}\sim +\partial_\l
\G^\l_\ab$, and the Levi--Civita symbol obeys $\e^{0...10} = -1$.
The gravitational constant $\kappa$, with dimension $[L]^{9/2}$
also appears explicitly in the topological $(P,T)-$ conserving
metric-independent Chern-Simon (CS) part of (\ref{Lagra}).

\noindent The propagators and vertices required for our
computations are obtained by expanding in powers of $\k$, with
$g_{\mu \nu}\equiv \eta_{\mu \nu}+\kappa h_{\mu \nu}$ and keeping
all contributions through order $\kappa^2$. The propagators, from
the quadratic part of the action, are well known. In harmonic (de
Donder) and Feynman gauges for gravity and the 3-form
respectively, \begin{equation} D^{\mu
\nu;\alpha\beta}(h)=\frac{1}{2} \left( \eta_{\alpha\mu
}\eta_{\beta\nu}+\eta_{\alpha\nu}\eta_{\beta\mu }- \frac{2}{9} \:
\eta_{\alpha\beta}\eta_{\mu \nu}\right) D_F \equiv G^{\mu
\nu;\alpha\beta}  D_F
\end{equation}
and%
\begin{equation} D^{\mu \nu\rho;\alpha\beta\sigma}(A)= \frac{1}{2}
\:\delta^{\mu \nu\rho}_{\alpha\beta\sigma}D_F \; ;
\end{equation}
$D_F$ is the scalar Feynman propagator and
$\delta^{\mu \nu\rho}_{\alpha\beta\sigma}$  is totally
antisymmetric in each triplet of indices.

There are three cubic vertices:

(a) Three gravitons $(h^3)$. Explicit use of this cumbersome
vertex can be avoided in dealing with the four-graviton
amplitudes, but not in computing the graviton-form ``Compton"
scattering. To minimize the complications, we write the vertex
already contracted with two on-shell polarization tensors, since
we will never need fewer contractions:%
\begin{eqnarray} &&2
V_{\mu\nu;\alpha\beta;\rho\sigma}(k^3,k^1,k^2)
\eps_1^\alpha\eps_1^\beta \eps_2^\rho\eps_2^\sigma =
     (\eps_1\cdot k_2)^2 \eps_2^{\mu}\eps_2^{\nu}
       -  (\eps_1\cdot k_2)(\eps_2\cdot k_1)
        \eps_1^{\mu}\eps_2^{\nu}
       -  (\eps_1\cdot k_2)(\eps_2\cdot k_1)
        \eps_2^{\mu}\eps_1^{\nu}\nonumber\\
       &&+  (\eps_2\cdot k_1)^2 \eps_1^{\mu}\eps_1^{\nu}
         +  (\eps_1\cdot \eps_2)(\eps_1\cdot k_2)(\eps_2\cdot k_1)
          \eta^{\mu\nu}
       -  (\eps_1\cdot k_2)(\eps_1\cdot \eps_2)\eps_2^\mu k_2^\nu
       -  (\eps_1\cdot k_2)(\eps_1\cdot \eps_2)\eps_2^\nu k_2^\mu\nonumber\\
&&       -  (\eps_1\cdot\eps_2)(\eps_2\cdot k_1)\eps_1^\mu k_1^\nu
       +  (\eps_1\cdot \eps_2)(\eps_1^{\mu}\eps_2^{\nu}
          +\eps_2^{\mu}\eps_1^{\nu}) k_1\cdot k_2
       -  (\eps_1\cdot\eps_2) (\eps_2\cdot k_1)\eps_1^\nu k_1^\mu
       +  (\eps_1\cdot\eps_2)^2 k_1^\mu k_1^\nu\nonumber\\
&&       + 1/2 (\eps_1\cdot\eps_2)^2 k_1^\mu k_2^\nu
       + 1/2 (\eps_1\cdot\eps_2)^2 k_1^\nu k_2^\mu
       +  (\eps_1\cdot\eps_2)^2 k_1^\mu k_2^\nu
       - 3/2 (\eps_1\cdot\eps_2)^2 \eta^{\mu\nu} k_1\cdot k_2
\end{eqnarray}
Here and throughout the polarization tensor $\e^\ab$ of a graviton
is represented as the product $\e^\a_i\e^\b_i$ of two polarization
vectors.

(b) Graviton-form  ($hFF$): this is the usual coupling between the
metric and the form's stress tensor. In coordinate space, %
\begin{eqnarray}%
 V^{gFF}_3= \kappa T_{\mu \nu} h^{\mu \nu}&=& \kappa
h^{\mu \nu} \left(F_{\mu \alpha\beta\rho}
F_{\nu}^{\sp\alpha\beta\rho} -\frac{1}{8} \: \eta_{\mu \nu}
F_{\alpha\beta\rho\sigma}F^{\alpha\beta\rho\sigma}\right)=\\ &=&
\kappa A_{\alpha\beta\rho} \partial_\mu \left (h^{\nu[\mu
}F_{\nu}^{\sp\alpha\beta\rho]}-\frac{h}{2} \: F^{\mu
\alpha\beta\rho}\right) \; .
\end{eqnarray}
Expressions (5,6) differ (onshell and for harmonic gauge) by an
integration by parts: the former is the more suitable in the
analysis of pure form scattering, the latter for graviton-form
Compton scattering. Note that both expressions simplify if we
choose a gauge where $h_\mn$ is traceless.

(c) Three formions ($AFF$): Due entirely to the (metric
independent) CS term in (\ref{Lagra}), it is usefully written as
\begin{equation}%
V_3^F=A_{\mu \nu\alpha}C^{\mu \nu\alpha}_F\ , \ \ \
C_F^{\rho\sigma\tau}\equiv\frac{2}{(12)^4} \:
\epsilon^{\rho\sigma\tau 1..8} F_{1..4} F_{5..8} \; .
\end{equation}
This vertex will produce a non-gravitational contribution to
4-formion scattering and will also be responsible for an unusual,
$F^3 R$ ``bremsstrahlung", amplitude.

Finally, to achieve gauge invariance, we must also include the
effects of two four-point contact $(\k^2)$ vertices. The first is
the local 4-graviton vertex; it will not be written out here, but
is needed for the 4-graviton amplitude calculation. The second is
the $hhFF$ vertex from expanding the $F^2$ kinetic term in (4); it
is necessary to insure gauge invariance in the graviton-formion
Compton process. Its form, in a gauge where the
graviton is traceless, is %
\begin{eqnarray}%
V^{hhFF} &  \equiv & \left. - \frac{1}{48} \: \d^2 \int \sqrt{-g}
\, F^2 / \delta g_\mn
 \d g_\ab  \right|_{g=\eta} h_\mn h_\ab
\nonumber \\%
& = & -\frac{\kappa^2}{12} \left[ h^\mu _{\sp\lambda}
h^{\nu\lambda} F_{\mu \alpha\beta\rho}
F_{\nu}^{\sp\alpha\beta\rho} -\frac{\kappa^2}{16} \: h_{\mu \nu}
h^{\mu \nu} F_{\alpha\beta\rho\sigma}F^{\alpha\beta\rho\sigma} +
\frac{\kappa^2}{2} \: h_{\mu \nu} h_{\alpha\beta} F^{\mu
\alpha}_{\sp\sp~\rho\sigma}F^{\nu\beta\rho\sigma}\right] \; .
\end{eqnarray}

\section{Amplitudes}

In this section we outline the explicit computation of (the
bosonic part of) the SUSY invariant amplitude and then construct
the corresponding local invariants. Before entering into details,
some general remarks are in order. In momentum space, the
non-locality in each scattering amplitude (due to the intermediate
denominator of the exchanged particle) is represented by a sum of
simple poles, in each of the Mandelstam variables ($s,t,u$),
corresponding to the three different possible channels in
four-particle scattering; this nonlocality is easily neutralized
by multiplying the final result by the symmetric polynomial
$stu~$.  Since multiplication in momentum space corresponds to
differentiation in coordinate space, it becomes necessary to
understand how these additional derivatives are to be spread.
Suppose that we can write the amplitude in the generic
``current-current" single pole form, as
follows \begin{equation}%
M=\phi^a_1(k_1) \phi_2^b (k_2)
V_{abm}(k_1,k_2)\,\frac{G^{mn}}{s}\, W_{mcd}(k_3,k_4)
\phi^c_3(k_3) \phi_4^d (k_4) + (stu) {\rm perm.}.
\end{equation}
Then, by multiplying by $stu$ and using the identity
\begin{equation} \label{Kmatrix} t u=-1/2 (\eta^{\mu \alpha}
\eta^{\nu\beta}+ \eta^{\nu\alpha} \eta^{\mu \beta}- \eta^{\mu \nu}
\eta^{\alpha\beta})k^1_\mu k^2_\nu k^3_\alpha k^4_\beta \equiv-1/2
K^{\mu \nu;\alpha\beta} k^1_\mu k^2_\nu k^3_\alpha k^4_\beta
\end{equation}
and its permutations, we can write%
\begin{equation}%
M=k^1_\mu \phi^a_1(k_1) k^2_\nu\phi_2^b (k_2) V_{abm}(k_1,k_2)
K^{\mu \nu;\alpha\beta} {G^{mn}}W_{mcd}(k_3,k_4)
k^3_\alpha\phi^c_3(k_3) k^4_\beta\phi_4^d (k_4) + (stu) {\rm
perm.} \; .
\end{equation}
In other words, if we Fourier-transform back to coordinate space,
the net effect of this procedure is to remove the pole and to add
a derivative to each of the four external fields. These new
derivatives are to be contracted according to the
$K^{\alpha\beta;\mu \nu}$ matrix defined in (\ref{Kmatrix}). If
the amplitude is already expressed as a product of gauge invariant
currents, this procedure produces an invariant that is the product
of two new dressed gauge-invariant currents. In the case of
gravitationally induced matter interactions, these currents behave
like counterparts of the Bel-Robinson (BR) tensors
\cite{Deser:1999me}.

The above ``dressing'' procedure leaves unaltered an amplitude's
transformation under global symmetries, such as the linearized
supersymmetry of interest:  We are just multiplying an invariant
by a numerical factor, the derivatives. While there will be some
exceptions in detail to application the above remarks, the final
local results achieved will be correct, {\it i.e.}, we have a
constructive procedure for transforming the guaranteed
symmetry-preserving but nonlocal amplitudes into equally invariant
(on-shell) local terms.

\subsection{$\mbox{\boldmath$R^4$}$: Graviton-Graviton Scattering}

We start with the 4-graviton amplitude $M^g_4$. The  graviton
exchange contributions stem from (a) contracting two $V^3_g$
vertices (4) in all three $(s,t,u)$ channels via an intermediate
graviton propagator (2), which provides a single denominator and
(b) the local 4-point vertex $V^g_4$. The resulting $M^g_4(h)$
will be  a non-local quartic polynomial in the Riemann (Weyl, on
linear shell) tensor, whose non-locality is removable by
$stu$-multiplication.  In D=4, most of the calculation can be
avoided because a straightforward implementation of supersymmetry
allows one to fix the amplitude completely up to normalization:
There are only two independent local scalar quartics in the Weyl
tensor and its dual, $^*\! R$ : the squares of Euler
($E_4\equiv~\!\! ^*\! R ^*\! R$) and Pontryagin ($P_4\equiv~\!\!
^*\! R R$) densities. Their relative coefficient can be determined
by exploiting the special property that ensures the
supersymmetrizability of the Einstein action, namely that it is,
at tree level, maximally helicity conserving
\cite{Christensen:1979qj}. This constrains the amplitude  to be
proportional to the combination $(E_4-P_4)(E_4+P_4)$. Remarkably,
this invariant is also, owing to identities peculiar to D=4,
expressible as the square of the (unique) BR tensor $B_{\mu
\nu\alpha\beta}=(R R+~\!^*\!\! R ^*\! R)_{\mu \nu\alpha\beta}$.
Unfortunately, $D=4$ is highly degenerate (see Appendix). In
generic dimension, which in this context means $D\ge 8$, the
number of invariants quartic in the Weyl tensor is seven and the
only condition given by the above constraint is obviously not
enough to fix the relative coefficients. Nevertheless it is still
sufficient to determine the amplitude completely by considering
configurations where the helicities of the gravitons belong to the
subspace defined by their four momenta.

A further step can be taken using a very different property, which
is not manifest from the GR action, having a string origin: The
4-graviton tree amplitude is proportional to the square of
``bleached" 4-gluon tree amplitudes, upon representing the
graviton polarization tensor as the product of two vectors; this
is implied by the  field theory limit of the KLT \cite{klt}
relations. This additional information in fact, determines the
amplitude completely, because maximal helicity conservation fixes
the (uncolored) 4-gluon amplitude (since there are only two
independent $F^4$ invariants in any $D$) and consequently the
gravity amplitude,
which is its square. The conclusion that the form of $M^g_4$ is%
$$
M^{g}_4\propto (s t u)^{-1} t_8^{\mu _1\cdots\mu _8}
t_8^{\nu_1\cdots\nu_8}
R_{\mu _1\mu _2\nu_1\nu_2}R_{\mu _3\mu _4\nu_3\nu_4}R_{\mu _5\mu _6\nu_5\nu_6}
R_{\mu _7\mu _8\nu_7\nu_8}  \eqno{\rm (12a)} %
$$%
follows from the gluon ``square root" (in this context,
$F_{\mu\nu}$ stands for the gluon field strength)%
$$
M^{gluon}_4\propto t_8^{\mu _1\cdots\mu _8} F_{\mu _1\mu _2}\cdots
F_{\mu _7\mu _8}=(F_{\mu \nu} F^{\mu \nu})^2 -4 F^{\mu _1\mu _2}
F_{\mu _2\mu _3}F^{\mu _3\mu _4}F_{\mu _4\mu _1} \; . \eqno{\rm (12b)}%
$$%

Alternatively one can follow the explicit calculational steps
spelled out at the beginning of this section. The algebra involved
is quite cumbersome, and benefits from a program for algebraic
manipulation. This analysis should obviously lead to  the same
result and indeed it does. Still, it must be performed, at least
for a particular set of helicities, in order to obtain the correct
normalization of the amplitudes.  For example, by choosing a
configuration such that $\epsilon_i\cdot k_j=0$  for all $i$ and
$j$, one finds that the overall coefficient of (12a) is fixed to
be $1/4$. The final result (12a) possesses the same tensorial
structure as the familiar superstring zero-slope limit correction
to D=10 N=2 supergravity, where the $t_8^{\mu \cdots \mu _8}$
symbol originates from the D=8 transverse subspace \cite{Schwarz},
as has also been noted in \cite{Sannan} which carried out the
direct 4-graviton calculations as well. This reflects the fact
that maximal supersymmetry implies a unique $R^4$ in all
dimensions. If we assume only $1/2$ of the maximal supersymmetry
in generic $D$ we find that there is room for two invariants, as
can be seen by looking, {\it e.g.} at the effective action of the
heterotic superstring where the analog of (12a) is accompanied by
another $R^4$ term\footnote{The number of supersymmetrizable $R^4$
combinations can be easily understood by means of the KLT
relations. Given two independent YM $F^4$, it is a straightforward
exercise to show that only three (combinations of) $R^4$ give rise
to an amplitude that factorizes into gauge invariant vector
amplitudes. Assuming ``$N=4$" supersymmetry requires just one of
the factors in the above product to be maximally helicity
conserving while assuming ``$N$=8" requires both, leaving just one
candidate. Obviously this reduces the number of invariants first
to two and then to one.}.

\setcounter{equation}{12} %
At this point it is quite easy to write down a combination of
local $R^4$ that represents (12a). In
terms of the basis of Appendix A, the Lagrangian is%
\begin{equation} \label{L4a} %
L^g_4=\frac{1}{4} \: I_4 - I_7,
\end{equation}
where we have dropped a  term proportional to the 8-dimensional
Euler density $(\e_8\e_8 R^4)$ that, being a total divergence to
leading order, does not affect the amplitude. In many respects,
the form (12a) for the contribution coming from the 4-graviton
amplitude, is a perfectly physical one. However, one might wonder
whether there is a formulation of the above Lagrangian in terms of
currents that encompasses both gravity and matter in a unified way
as in fact occurs in {\it e.g.} $N=2,D=4$ supergravity
\cite{deser}. This might also lead to some understanding of higher
spin  SUSY multiplets. Using the quartic basis expansion, one may
rewrite $L^g_4$ in various ways involving any of the BR currents
of the Appendix and a closed 4-form $P_{\alpha\beta\mu \nu}=1/4
R^{ab}_{\sp\sp[\mu \nu} R_{\alpha\beta]a b}$. For example if we
choose the BR tensor%
$$ B_{\mu \nu\alpha\beta}\equiv%
[R_{\mu\rho\alpha\sigma} R^{\ ~\rho\ ~\sigma}_{\nu\ ~\beta\ } +
(\nu\mu)]
-\frac{1}{2} \: g_{\mu \nu} R_{\alpha\rho\sigma\tau} R_{\beta}^{\
~\rho\sigma\tau}- \frac{1}{2} \: g_{\alpha\beta} R_{\mu
\rho\sigma\tau} R_{\nu}^{\ ~\rho\sigma\tau} +\frac{1}{8} \: g_{\mu
\nu} g_{\alpha\beta}
R_{\lambda\rho\sigma\tau}R^{\lambda\rho\sigma\tau}, $$
we can write %
\begin{equation}%
L^g_4=48 \kappa^2\left[ 2 B_{\mu \nu\alpha\beta}B^{\mu
\alpha\nu\beta} -B_{\mu \nu\alpha\beta}B^{\mu \nu\alpha\beta}+6
B_{\mu \rho\alpha}^{\ \ \ \ \rho}B^{\mu \sigma\alpha}_{\ \ \ \
\sigma} -\frac{15}{49} \: (B^{\mu \nu}_{\ \ ~\mu \nu})^2 + P_{\mu
\nu\alpha\beta}P^{\mu \nu\alpha\beta}  \right].
\end{equation}
Due to the larger number of allowed invariants and of helicities
in D=11, this representation does not seems to share the elegance
and power of the four dimensional one. Still, it is remarkably
compact.

\subsection{$\mbox{\boldmath $F^4$}$: Formion-Formion Scattering}

We turn now to pure formion scattering. This amplitude is quite
simple to investigate because it must be manifestly (form) gauge
invariant: the three-form  potential $A$ only appears in the
operative vertices (5,7) through  its curvature $F$; the relevant
currents are in fact the CS $C^F_{\mu \nu\alpha}$ and the stress
tensor $T^F_{\mu \nu}$. The interactions are mediated respectively
by the formion and the graviton. Therefore the amplitude is
already organized in terms of gauge invariant
currents; indeed we have, in terms of $T_F, \: C_F$ of (5,7),%
\begin{equation}%
 M^{grav-med.}_{F4}=4 \left(\frac{\kappa}{12}\right)^2
\left ( T^{\alpha\beta}_F(k_1,k_2)\frac{1}{s} \:
G_{\alpha\beta;\mu \nu} T^{\mu \nu}_F(k_3,k_4)+{\rm perm.}\right).
\end{equation}
and %
\begin{equation}%
M^{form-med.}_{F4} =-\frac{1}{12} \left(\frac{\kappa}{24}\right)^2
\left ( C^{F~\alpha\beta\rho}(k_1,k_2)\frac{1}{s} \:
C^F_{\alpha\beta\rho}(k_3,k_4)+{\rm perm.}\right),
\end{equation}
where ``perm" stands for permutation of the four external
particles. The sum of (15) and (16) agrees with a recent
calculation of formion scattering from a quite different
starting-point \cite{Plefka:1998xu}. We must now multiply our
total $M_{F4}$ by $stu$ and see how the derivatives spread. Using
the simple rule stated at the beginning of this section, we
recognize immediately that there is an economical way of
organizing  $L^F_4$ in terms of matter BR and of the corresponding
$C^F$ extensions. In fact if we define
\begin{eqnarray} \label{BF} &&B^F_{\mu \nu\alpha\beta} \equiv
\partial_{\alpha} F_\mu
\partial_{\beta} F_{\nu}+
\partial_{\beta} F_\mu
\partial_{\alpha} F_{\nu}-
\frac{1}{4} \: \eta_{\mu \nu} \partial_{\alpha} F
\partial_{\beta} F, \ \ \ \partial^\mu B^F_{\mu \nu\alpha\beta}=0,\\
&&C^F_{\rho\sigma\tau;\alpha\beta} \equiv \frac{1}{(24)^2} \:
\epsilon_{\rho\sigma\tau\mu _1\cdots\mu _8}
\partial_\alpha F^{\mu _1\cdots\mu _4} \partial_\beta
F^{\mu _5\cdots\mu _8},\ \ \ \partial^\rho C^F_{\rho\sigma\tau;\alpha\beta}=0.
\end{eqnarray}
where implicit indices are summed in the obvious way, then%
\begin{equation} \label{BF1}%
 L^F_4=\frac{\kappa^2}{36} \: B^F_{\mu
\nu\alpha\beta} B^{F}_{\mu _1\nu_1\alpha_1\beta_1} G^{\mu \mu
_1;\nu_1\nu}K^{\alpha\alpha_1;\beta_1\beta}-\frac{\kappa^2}{12} \:
C^F_{\mu \nu\rho;\alpha\beta}C^{F\mu
\nu\rho}_{~~~~~~\alpha_1\beta_1} K^{\alpha\alpha_1;\beta_1\beta}.
\end{equation}
Reflecting its simple ``current-current" origin, the pure matter
sector has a natural (if perhaps not unique) expression in terms
of currents. There is also a basis of scalars quartic in $F$; we
have not used it here, but it is tabulated in the Appendix.

\subsection{$\mbox{\boldmath $F^3 R$}$: Topological Bremsstrahlung}

Here there is just one diagram, namely the emission of a graviton
described by the stress tensor vertex (5,6), from any of the 3
formion lines emanating out of the CS vertex (7). The analysis of
this amplitude follows the lines of the previous section. While it
is not manifestly (gravitationally) gauge-invariant, its
invariance
can be verified using the following local D=11 identity%
\begin{equation} \label{cip}%
d(A\wedge F\wedge F\wedge F) \equiv F_{\alpha\mu _1\mu _2\mu _3}
F_{\mu _4\cdots\mu _7}F_{\mu _8\cdots\mu _{11}} \epsilon^{\mu
_1\cdots\mu _{11}} \equiv 0.
\end{equation}
This identity enables us to write the amplitude schematically in
the form%
\begin{equation} \label{cip1}%
M^{hF^3}_4= h^{\mu \nu} \left(F_{\mu \alpha\beta\rho}
G_{\nu}^{\sp\alpha\beta\rho} -\frac{1}{8}\: \eta_{\mu \nu}
G_{\alpha\beta\rho\sigma}F^{\alpha\beta\rho\sigma}\right) +{\rm
perm}
\end{equation}
where ``perm" symmetrizes the 3 formions; $G_{\mu \nu\alpha\beta}$
is the effective field strength (obeying the $F$ equation of
motion) constructed out of the ``connection" defined by
${(\triangle)^{-1} }C_F^{\alpha\beta\rho}$ with the $C_F$ of (7).
Then the gauge invariance of the amplitude is equivalent to the
conservation of the ``energy momentum tensor" effectively defined
in (\ref{cip1}). Next we again multiply derivatives according to
the rule given at the start of this section. Turning the
$h_{\mu\nu}$ in (\ref{cip1}) into a Riemann tensor takes some
patience and a certain number of integration by parts, however.
The final result is %
\begin{eqnarray} %
&& L_4^{FFF g}= (s t u) M_4^{FFFg}= -\frac{\kappa^2}{3} \:
C^F_{\mu \nu\rho;\alpha\beta} C^{RF\mu
\nu\rho}_{~~~~~~~\alpha_1\beta_1}K^{\alpha\alpha_1;\beta_1\beta},\\
&&C^{RF}_{\mu \nu\rho;\alpha\beta} \equiv 4\partial_\lambda\left(
R^{\sigma\  [\lambda }_{\ (\alpha\ ~\beta)} F_{\sigma}^{\ \mu
\nu\rho]}\right ) -\frac{2}{3}\: R^{~\sigma\  ~\lambda }_{\
~(\alpha\ ~\beta)}\partial_\lambda F_{\sigma}^{\ \mu \nu\rho} \; .
\end{eqnarray}
To prove this result, we used the following generalization of the
identity (\ref{cip})%
\begin{equation} \label{cip3}%
\partial_\alpha\partial_\beta F_{\alpha\mu _1\mu _2\mu _3}
\partial^\alpha F_{\mu _4\cdots\mu _7}\partial^\beta F_{\mu _8\cdots\mu _{11}}
\epsilon^{\mu _1\cdots\mu _{11}}\equiv 0.
\end{equation}
While it is clear that a ``$C^{RF}$ current" must exist since
$C^F$ factorizes the amplitude, (23) is not unique and we claim no
special significance for it.

\subsection{$\mbox{\boldmath $R^2F^2$}$: Compton Scattering}

The most complicated amplitude is that for graviton-formion
scattering. It involves two classes of diagrams. The first
consists of the $T^F_\mn$ stress tensor turning into two gravitons
via graviton exchange between the vertices $h^\mn T^F_\mn$ of (6)
and the $h\pa h\pa h$ of (4) along with the mixed quartic contact
term (8) required to preserve gauge invariance. The second set is
more Compton-like: the gravitons scatter off formion lines, via
two $T^F_\mn$ currents through virtual formion exchange (in direct
as well as crossing versions). The schematic expression for the
total amplitude should look like $M^{ggFF}_4 \sim \k^2 R^2 F^2$ up
to derivatives and the exchange pole. [There is no simple D=4
reduction available here since a 4-form is a constant in D=4.]  To
perform the detailed calculations it proved useful to employ the
program FORM \cite{FORM}.

As yet we can only give the amplitude in semi-final form, before
the graviton polarizations have been converted into curvatures,
but with the formions entirely expressed in terms of their field
strengths. The eventual ``FFRR" form is guaranteed by the
(verified) invariance of M under graviton gauge transformations.
The amplitude, before ($stu$) multiplication, reads%
\begin{eqnarray}
     &&\!\!\!\!\!M^{ggFF}_4=\nonumber\\
     &&\!\!\!\!\!
         \frac{1}{6 s}\Biggl (
          F_{12}^{\mu_1\nu_1}\eps_{1\mu_1}\eps_{2\nu_1}\eps_2.p_2 \eps_1.p_1
          - 3 F_{12}^{\mu_1\mu_2\nu_1\nu_2}
            \eps_{2\mu_1}\eps_{1\mu_2} k_{2\nu_1}\eps_{2\nu_2} \eps_1.p_1
          - 3 F_{12}^{\mu_1\mu_2\nu_1\nu_2}
              k_{1\mu_1}\eps_{1\mu_2}\eps_{1\nu_1}\eps_{2\nu_2}
             \eps_2.p_2
\nonumber\\
         &&- 6 F_{12}^{\mu_1\mu_2\mu_3\nu_1\nu_2\nu_3}
            k_{1\mu_1} \eps_{2\mu_2}\eps_{1\mu_3} k_{2\nu_1}
            \eps_{1\nu_2} \eps_{2\nu_3}
          + 3 F_{12}^{\mu_1\mu_2\nu_1\nu_2}
             k_{1\mu_1}\eps_{1\mu_2} k_{2\nu_1}\eps_{2\nu_2}
 \eps_1.\eps_2
          \Biggr)     +\nonumber\\
  &&\!\!\!\!\!\frac{1}{6 u}\Biggl (
          F_{12}^{\mu_1\nu_1}\eps_{2\mu_1}\eps_{1\nu_1}\eps_1.p_2 \eps_2.p_1
          - 3 F_{12}^{\mu_1\mu_2\nu_1\nu_2}
            \eps_{1\mu_1}\eps_{2\mu_2} k_{1\nu_1}\eps_{1\nu_2} \eps_2.p_1
- 3 F_{12}^{\mu_1\mu_2\nu_1\nu_2}
              k_{2\mu_1}\eps_{2\mu_2}\eps_{2\nu_1}\eps_{1\nu_2}
             \eps_1.p_2
\nonumber\\
         &&- 6 F_{12}^{\mu_1\mu_2\mu_3\nu_1\nu_2\nu_3}
            k_{2\mu_1} \eps_{1\mu_2}\eps_{2\mu_3} k_{1\nu_1}
            \eps_{2\nu_2} \eps_{1\nu_3}
            + 3 F_{12}^{\mu_1\mu_2\nu_1\nu_2}
             k_{2\mu_1}\eps_{2\mu_2} k_{1\nu_1}\eps_{1\nu_2}
 \eps_1.\eps_2
          \Biggr)+     \nonumber\\
&&\!\!\!\!\! \frac{1}{6 t}  \Biggl (
          F_{12}^{\mu_1\nu_1}
            \eps_{2\mu_1} \eps_{1\nu_1}\eps_2.k_1 \eps_1.k_2
          + F_{12}^{\mu_1\nu_1}\eps_{1\mu_1}\eps_{2\nu_1}
           \eps_2.k_1 \eps_1.k_2
          - \frac{1}{2}F_{12}^{\mu_1\nu_1} k_{1\mu_1}k_{2\nu_1}
      (\eps_1. \eps_2)^2
        \nonumber\\
      &&- \frac{1}{2}F_{12}^{\mu_1\nu_1}k_{2\mu_1}k_{1\nu_1}(\eps_1. \eps_2)^2
            - F_{12~\mu_1\nu_1}{\cal F}_{1}^{\mu_1\alpha_1} \eps_{2\alpha_1}
                    {\cal F}_{1}^{\nu_1\alpha_1}\eps_{2\alpha_1} -
        F_{12~\mu_1\nu_1}{\cal F}_{2}^{\mu_1\alpha_1} \eps_{1\alpha_1}
                     {\cal F}_{2}^{\nu_1\alpha_1}\eps_{1\alpha_1}
\Biggr ) \nonumber\\
       &&\!\!\!\!\!+ \frac{1}{12} \Biggl
     (F_{12}^{\mu_1\nu_1}\eps_{2\mu_1}\eps_{1\nu_1}
                     \eps_2.\eps_1
          + 6 F_{12}^{\mu_1\mu_2\nu_1\nu_2}\eps_{1\mu_1}
     \eps_{2\mu_2}\eps_{1\nu_1}\eps_{2\nu_2}
          +       F_{12}^{\mu_1\nu_1}\eps_{1\mu_1}\eps_{2\nu_1} \eps_1.\eps_2
          +\frac{1}{8} F_{12} (\eps_1.\eps_2)^2 \Biggr) \; .
\end{eqnarray}
The last, local, term includes the 4-point vertex (8) as well as
local contributions from the other graphs. The notation is as
follows: $k_i,p_i$ denote respectively the graviton and formion
momenta, $F_{12}^{\mu_1,\cdots \mu_i\nu_1,\cdots \nu_i}$ is the
product of the field strengths of formions 1 and 2, with the last
$4-i$ indices contracted, while ${\cal F}_i^{\mu_1\mu_2}$ stands
for the invariant combinations
$k_i^{\mu_1}\eps_i^{\mu_2}-k_1^{\mu_2}\eps_i^{\mu_1}$. As in Sec.
2, the polarization tensor of each graviton is represented as the
product of two polarization vectors, $\epsilon_i$. With these
conventions, the amplitude is symmetric under $s-u$ interchange,
corresponding to interchange of the (1-2) gravitons, while the
$1/t$ term is then separately invariant under (1-2).

In summary, the set of scattering amplitudes (14,19,22,25)
displayed in this section represents the bosonic part of the
advertised linear 4-point SUSY invariant.

\section{Local Invariants and the Renormalization Problem}

 In the previous section,
we first constructed and then localized the (bosonic) four-point
tree amplitudes to obtain the bosonic part of a linearized SUSY
invariant quartic in the field strengths $(F,R)$. Here we discuss
some consequences of this invariant's existence on the issue of
renormalizability of $D=11$ SUGRA. In this connection a brief
review of the general SUGRA divergence problem as it applies to
D=11 may be useful. For clarity, we work in the framework of
dimensional regularization, in which only logarithmic divergences
appear and consequently the local counterterm must have dimension
zero (including  dimensions of the coupling constants in the loop
expansion); the generic gravitational loop expansion proceeds in
powers of $\kappa^2$ (we will separately discuss the effect of the
additional appearance of $\kappa$ in the CS vertex).  It should
also be stated (in connection with another $\k^2$ counting) that
while the present discussion really proceeds at lowest order in an
expression about flat space, with linearized curvatures, etc, the
``covariantly dressed" quantities enter through including
additional graviton lines at each graviton vertex; this will not
alter the divergence countings, although it can be extremely
complicated to achieve. Indeed, the same can be said of the whole
process of reaching the fully locally SUSY invariant version of
our 4-point amplitudes: it must exist just because it comes from
the underlying action (1), as the physical expression of
scattering among asymptotically defined states, though that does
not make the perturbative resummation very obvious!

At one loop (omitting the overall ``infinite" $1/\epsilon$
factor), the counter-action would be $\triangle I_1\sim \kappa^0
\int dx^{11} \triangle L_1$. But there is no candidate $\triangle
L_1$ of dimension $11$, since odd dimension cannot be achieved  by
a purely gravitational $\triangle L_1$. [``Gravitational'' $\sim
\epsilon\Gamma R^4$ or ``form-gravitational'' $\sim \epsilon A
R^4$ (respectively parity odd and even) CS-like\footnote{In this
connection we also note that the presence of the Levi-Civita
tensor usually does not invalidate the use of dimensional
regularization (or reduction) schemes to the order we need. In any
case our conclusions would also apply, in  a more complicated way,
in other regularization schemes that preserve SUSY.} terms
\cite{liu} cannot arise perturbatively {\it i.e.}, with integer
powers of $\k$.]  Possible invariants involving odd powers of
$\kappa$ arising from the CS vertex also cannot give rise to
$1-$loop diagrams.  These candidates, consisting of a polygon
(triangle or higher) with form/graviton segments and appropriate
emerging external bosons at its vertices, have as simplest example
a form triangle with three external $F-$lines $\sim\kappa^3\int
d^{11} x
\partial^9 \epsilon  A F F  $. However, this odd number of derivatives
clearly cannot yield a scalar. The same counting also excludes the
one-loop polygon's  gravitational or form extensions such as $F^2
R$, $F R^2$ or even $F^3 R$ at this $\kappa^3$ level.

At two loops, $\triangle I_2\sim\kappa^2\int d^{11}x \triangle
L_2$, so that $\triangle L_2\sim [L]^{-20}$ which can be achieved
(to lowest relevant, 4$^{\rm th}$, order in external lines) by
(for the pure graviton contribution) $\triangle L_2 \sim
\partial^{12} R^4$, where $\partial^{12}$ means twelve explicit
derivatives spread among the 4 curvatures. There are no relevant
$2-$point $\sim
\partial^{16} R^2 $ or $3-$point $\sim \partial^{14} R^3 $ terms
because the $R^2$ can be field-redefined away into the Einstein
action in its leading part (to $h^2$ order, $E_4$ is a total
divergence in any dimension) while $R^3$ cannot appear by SUSY.
This latter fact was first demonstrated in  D=4 but must therefore
also apply in higher D simply by a direct dimensional reduction
argument. Thus the terms we need are, for their 4-graviton part,
$L^g_4$ of (5) with twelve explicit derivatives. The companions of
$L_4^g$ in $L^{tot}_4$  will simply appear with the same number of
derivatives. It is easy to see that the additional $\partial^{12}$
can be inserted without spoiling $SUSY$; indeed they appear as
naturally as did multiplication by $s t u $ in  localizing the
$M_4$ to $L_4$: for example, $\partial^{12}$ might become, in
momentum space language, a combination of $(s^6+t^6+u^6)$ and
$(stu)^2$, spread according to rules similar to those presented in
the text. This establishes the structure of the $4-$point local
counterterm candidate we are considering.

Before the present construction of the complete counterterm was
completed, the actual coefficient of its 4-graviton part was
computed \cite{bern} by a combination of string-inspired and
unitarity techniques.  The structure of infinities in the
four-graviton sector for all maximal supergravities up to two
loops was extensively studied there, and conjectures on higher
loops were presented as well. [Very recently, a parallel analysis
of type I supergravities has been carried out in \cite{dunbar}].
Here, for completeness we state the methods and relevant final
results of \cite{bern}: Begin by computing the tree supergravity
amplitudes by means of the KLT relations. Next, use these tree
amplitudes as input for the cutting rules to obtain the analytic
structure of the one-loop amplitudes at any $D$. This information,
because of the high degree of supersymmetry, is enough to
reconstruct the one-loop amplitudes. Now iterate the procedure and
go to two loops. [What makes the procedure quite cumbersome beyond
two loops is the increasing number of $n-$particles cuts that one
has to examine to reconstruct the amplitude\footnote{Remarkably,
the two-particle cut can be iterated to an arbitrary number of
loops, because ``N=8" supersymmetry guarantees very simple
iterative rules for gluing the amplitudes. One essentially always
reproduces the tree level's tensorial structure.}.] Finally
compute the eventual divergences; in $D=11$ as we saw on general
ground there is no 4-point one-loop divergence, while at two loops
the calculation
yields the explicit infinite result%
\begin{equation} %
\triangle I_2 (g)\vert_{\rm pole}
 =  \left( {\kappa \over 2}\right)^6 {1\over 48\e\ (4\pi)^{11}} {\pi\over 5791500}
  \Bigl( 438 (s^6+t^6+u^6) - 53 s^2 t^2 u^2 \Bigr) \,
   (stu M_4^{\rm g ~tree}) \, ,
\end{equation}
for the local 4-graviton divergence, dimensionally regularized to
D=11-2$\epsilon$, with $(stu) M^{\rm g~tree}_4 \equiv L^g$ of
(12a--14). The results of Sec. 3 then embody the extension of (26)
to the complete bosonic sector counterterm.

\section{Summary}

We have succeeded in constructing explicitly the tree level
nonlocal 4-point scattering amplitudes involving the two bosons of
D=11 SUGRA, namely the graviton and formion, as well as obtaining
the corresponding local invariant in a SUSY-preserving way.
Extending the result to the rest of the amplitude, involving two
or four gravitinos, is not that difficult in terms of the
techniques employed here \cite{waldron}: the gravitino primarily
interacts with the graviton through its stress tensor $\sim k^\mn
T_\mn (\psi )$, and with the formion through a simple (nonminimal)
coupling term $\sim (\bar{\psi} \Gamma \psi F)$.  The
(complicated) 4-fermion contact terms are needed, but only for the
4-fermion part of the amplitude, where they insure the SUSY
invariance, not for the 2-gravitino to 2-boson amplitudes. In any
case the bosonic part alone, if SUSY-transformed, will provide a
complete linearized SUSY invariant. In addition to its intrinsic
interest as a example of a ``physical" process in D=11 SUGRA, the
result was of primary interest to us as confirmation of the
existence of an invariant that, (in its localized version) has the
dimension of a candidate counterterm for (dimensionally
regularized) 2-loop infinities. Indeed, its 4-graviton part agreed
completely with the coefficient of the 2-loop infinity recently
calculated in that sector in \cite{bern}, while its 4-formion part
agreed with a very different matrix-theory motivated scattering
calculation \cite{Plefka:1998xu}.  The existence of infinities in
this ultimate local SUGRA model, while not unexpected from a
purely power counting field theoretical point of view, is
important in showing that no hidden symmetry rescues this most
unique theory. Of course such a putative symmetry could still
suppress all higher loop infinities beyond a certain order, but
this seems unlikely given the concrete result of \cite{bern},
together with the obvious constructibility of higher order
candidate counterterms {\it e.g.}, using the scattering approach.
We can at least conclude that the case for underlying finite
extended (M-)theories is thereby strengthened.  In this
connection, we emphasize that the invariant found here has a
further interest as another example (see also \cite{green,russo})
of possible local corrections to M-theory whose leading term is
presumably the action (1). This might teach us something about
this underlying model, just as the corrections to the Einstein
action in slope expansion of the various D=10 superstring models
could be understood from the latters' properties; persistence in
D=11 of the ``$t_8t_8$" D=10 string theory hallmark is perhaps one
first hint about the M-string connection.

{\bf Acknowledgements}: We are grateful to many colleagues,
especially to Z.\ Bern, L.\ Dixon, and D.\ Dunbar for discussions
of their work and ours, as well as to A.\ Waldron, who also
introduced us to the world of FORM and to J.\ Franklin and S.\
Fulling for help with algebraic aspects of BR-ology.    This work
was supported by the National Science Foundation under grant
PHY99-73935.

\newpage

\noindent{\large\bf Appendix: Quartic Curvature Invariant Basis}

\renewcommand{\theequation}{A\arabic{equation}}
\setcounter{equation}{0}

We briefly tabulate (for the Ricci-flat geometries of interest) a
basis in the space of scalars quartic in  curvature, with
attention to the special case D=4; details can be found in
\cite{Deser:1999me,fulling}. A similar expansion of quartics
scalar in the 4-form fields is also appended.

In generic dimension, which turns out to be D$\geq$8, the basis
consists of 7 elements. A suitable choice \cite{fulling} is given
by (we retain the letter $R$ to denote
on-shell Riemann, that is, Weyl tensors):%
\begin{equation} \begin{array}{lll} I_1=(R_{\alpha\beta\rho
\sigma} R^{\alpha\beta\rho\sigma})^2, &
I_2=R^{\alpha\beta\rho\sigma}R_{\alpha\beta\rho}
^{\phantom{\alpha\beta\rho}\lambda}
R^{\mu\nu\omega}_{\ph{\mu\nu\omega}\sigma}R_{\mu\nu\omega\lambda},
& I_3=R^{\alpha\beta\rho\sigma}
R_{\alpha\beta}^{\ph{\alpha\beta}\lambda\mu}
R_{\lambda\mu}^{\ph{\lambda\mu}\nu\omega}R_{\rho\sigma\nu\omega},\\
I_4=R^{\alpha\beta\rho\sigma}
R_{\alpha\beta}^{\ph{\alpha\beta}\lambda\mu
}R_{\rho\lambda}^{\ph{\rho\lambda} \nu\omega}
R_{\sigma\mu\nu\omega}, &
I_5=R^{\alpha\beta\rho\sigma}R_{\alpha\beta}^{\phantom{\alpha\beta}\lambda\mu}
R^{\ph{\mu}\nu\ph{\omega}\omega}_{\rho\ph{\mu}\lambda\ph{\mu}}R_{\sigma\nu\mu\
\omega}, & I_6=R^{\alpha\beta\rho\sigma}
R_{\alpha\ph{p}\rho}^{\ph{p}\lambda\ph{p}\mu}R_{\lambda\sp
\mu}^{\sp \nu\sp \omega}R_{\beta\nu\sigma\omega},\\
I_7=R^{\alpha\beta\rho\sigma}
R_{\alpha\ph{p}\rho}^{\ph{p}\lambda\ph{p}\mu} R_{\lambda\sp
\beta}^{\sp \nu\sp \omega}R_{\mu\nu\sigma\omega}. & &
\end{array}
\end{equation}
Since we are  actually interested only in actions (rather than
local scalars) we are free to discard any combinations of the
$I_n$ that produce a total divergence at the linearized level. The
Euler density, \begin{equation} \label{E8} E_8= I_1-16 I_2+2
I_3+16 I_4-32 I_5+16 I_6-32 I_7,
\end{equation}
possesses this property in every dimension and thus the
combination (\ref{E8}) can be considered as effectively vanishing:
for our purpose there are then only $6$ independent invariants.
[$E_8$ could be detected in amplitudes with more than four
gravitons.]

In $D$=4 the number of independent invariant is further reduced to
just 2. The relation connecting the different $I_i$ can be shown
in many different ways. Here, to be self-contained, we will
demonstrate them by exploiting some ``accidental" symmetries of
the Bel-Robinson tensor $B_{\mu\nu\alpha\beta}$: Upon expanding
the product of Levi--Civita symbols implicit in the two dual
curvatures of%
\begin{equation} \label{BR1} B_{\mu
\nu\alpha\beta}=R^{\rho\sp\sigma}_{\sp\mu \sp\alpha}
R_{\rho\nu\sigma\beta}+ \dual{R}^{\rho\sp\sigma}_{\sp\mu
\sp\alpha} \dual{R}_{\rho\nu\sigma\beta} \; ,
\end{equation}
one obtains the different form %
\begin{equation} \label{BR2} B_{\mu
\nu\alpha\beta}=R^{\rho\sp\sigma}_{\sp\mu \sp\alpha}
R_{\rho\nu\sigma\beta}+R^{\rho\sp\sigma}_{\sp\mu \sp\beta}
R_{\rho\nu\sigma\alpha}-\frac{1}{2} \: g_{\mu \nu}
R_{\alpha}^{\sp\rho\sigma\tau} R_{\beta\rho\sigma\tau}
\end{equation}
of the same object. We thereby easily recover the two D=4
identities \begin{equation} \label{Id1}
R_{\alpha}^{\sp\rho\sigma\tau} R_{\beta\rho\sigma\tau}-\frac{1}{4}
\: g_{\alpha\beta} R^{\mu \rho\sigma\tau} R_{\mu \rho\sigma\tau}=0
\; ,
\end{equation}
\begin{equation} \label{Id2} \frac{1}{2}R_{\mu
\alpha}^{\sp\sp\rho\sigma} R_{\nu\beta\rho\sigma} +R^{\rho\sp
\sigma\sp }_{\sp \mu \sp \beta} R_{\rho\nu\sigma\alpha}-R^{\rho\sp
\sigma\sp }_{\sp \nu\sp \mu }
R_{\rho\alpha\sigma\beta}-\frac{1}{8} \: (g_{\mu \nu}
g_{\alpha\beta} - g_{\mu \beta}
g_{\nu\alpha})R_{\lambda\rho\sigma\tau}
R^{\lambda\rho\sigma\tau}=0 \; .
\end{equation}
The first identity follows from tracelessness of (\ref{BR1}) in
its first index pair, a property manifest for (\ref{BR2}).  The
second follows by exploiting the total symmetry of (\ref{BR2}).
Both facts are implicit in the (\ref{BR1}) definition.  Then
(\ref{Id1}) implies
\begin{equation} \label{rel1} %
I_1= 4 I_2.
\end{equation}
The scalar identities coming from (\ref{Id2}) are obtained by
multiplying it by all possible independent 4-index tensors: using
$R^{\rho \mu \sigma \nu} R_{\rho\sp \sigma}^{\sp\sp \alpha\sp\sp
\beta}$, we find
\begin{equation} \label{rel2}%
 I_7-I_6+\frac{1}{2} \: I_5-\frac{1}{4} \:
I_4+\frac{1}{8} \: I_1=0 \; ,
\end{equation}
while  $R^{\rho \mu \sigma \nu} R_{\rho \sp \sigma}^{\sp\sp
\beta\sp\sp \alpha}$ yields
\begin{equation} \label{rel3}%
I_7-I_6+\frac{1}{8} \: I_3+\frac{1}{16}\: I_1=0.
\end{equation}
Using the vanishing of the Euler combination and the relations
(\ref{rel1}), (\ref{rel2}) and (\ref{rel3}), we can, for example,
determine all the invariants in terms of $I_1, I_3$ and $I_5$
\begin{equation} I_7=\frac{I_1}{8}-\frac{I_3}{4},\ \ \
I_6=\frac{3}{16} \ I_1 -\frac{I_3}{8},\ \ \ \
I_4=\frac{I_1}{4}-\frac{I_3}{2}+ 2 I_5, \ \ \ \ I_2=\frac{I_1}{4}.
\end{equation}
Finally, to show that $I_5$ vanishes identically, one utilizes the
vanishing in $D$=4 of any expression antisymmetric in 5 indices;
more specifically antisymmetrizing the 5 lower (or upper) indices
$(pqtuw)$ in the definition of $I_5$ in (27) yields (after some
algebra) the value 4! $I_5$.

Although we have not explicitly used them in text, similar
(off-shell) bases also exist for our 4-forms.  For D$\geq$8, there
are 4 independent combinations,
\begin{eqnarray}%
f_1 & \equiv & (F^2)^2 \; , \;\;\;\; f_2 \; \equiv \; (F^\mu
F^\nu ) (F_\mu F_\nu ) \nonumber \\%
f_3 & \equiv & F_{AB} F^{BC} F_{CD} F^{DA} \nonumber \\%
f_4 & \equiv & F_{A\mu\nu} F^A_{~\alpha\beta} F_B^{~\mu\alpha}
F^{B\nu\beta}
\end{eqnarray}
where the missing indices in $f_1$, $f_2$ are internally traced in
each pair, while the capital indices in $f_3$, $f_4$ are shorthand
for an index pair.  Thus, each pair in $f_1$ has no open indices,
in $f_2$ there are two, and $f_3$, $f_4$ have 4 open indices per
pair, traced in the two possible independent ways.

\end{document}